\documentclass{appolb}
\usepackage[latin9]{inputenc}
\usepackage{amsthm}
\usepackage{amsmath}
\usepackage{graphicx}
\usepackage{esint}

\makeatletter

\providecommand{\tabularnewline}{\\}

\numberwithin{equation}{section}
\numberwithin{figure}{section}

\usepackage{graphicx}

\makeatother

\begin{document}
\title{\mbox{$X(3872)$ electromagnetic decay in a coupled-channel
model\thanks{Presented by M.~Cardoso at the Workshop ``EEF70'', Coimbra,
Portugal, September 1--5, 2014}}}
\author{Marco Cardoso\address{CFTP, Instituto Superior T\'{e}cnico,
Universidade de Lisboa, Lisbon, Portugal}
\\[5mm]
George Rupp\address{CFIF, Instituto Superior T\'{e}cnico,
Universidade de Lisboa, Lisbon, Portugal}
\\[5mm]
Eef van Beveren\address{CFC, Departamento de F\'{\i}sica,
Universidade de Coimbra, Coimbra, Portugal}
}

\maketitle

\begin{abstract}
A multichannel Schr\"{o}dinger equation with both quark-antiquark and
meson-meson components, using a harmonic-oscillator potential
for $q\bar{q}$ confinement and a delta-shell string-breaking potential for
decay,
is applied to the axial-vecor $X(3872)$ and lowest vector charmonia. The model
parameters are fitted to the experimental values of the masses of
the $X(3872)$, $J/\psi$ and $\psi(2S)$. The wave functions of these
states are computed and then used to calculate the electromagnetic
decay widths of the $X(3872)$ into $J/\psi\gamma$ and $\psi(2S)\gamma$. 
\end{abstract}

\section{Introduction}

The $X(3872)$ was discovered in 2003 by the Belle Collaboration
\cite{Choi:2003ue}, and later confirmed in CDF \cite{Acosta:2003zx} and D0
\cite{Abazov:2004kp} experiments. Its PDG\cite{Agashe:2014kda}
mass and width are now
$M_{X}=3871.69\pm0.17\,\mbox{MeV}$ and $\Gamma_{X}<1.2\,\mbox{MeV}$,
respectively.
According to experiment it has quantum numbers
$J^{PC}=1^{++}$\cite{Aaij:2013zoa}
and $I^{G}=0^{+}$ \cite{Aubert:2004zr,Choi:2011fc}. The $X(3872)$ seems to be
difficult to describe as a simple $c\bar{c}$ state.

Its main decays are into $\rho^{0}J/\psi$, $\omega J/\psi$
and $DD\pi$, with the latter final state resulting mainly from an intermediate
$DD^{*}$ channel. The first two channels are OZI forbidden and the decay into
$\rho^{0}J/\psi$ also violates isospin conservation. Both are therefore
highly suppressed. As the mass is below the $DD^{*}$ thresholds
($E_{D^{0}D^{0*}}=3871.84$ MeV and $E_{D^{\pm}D^{\mp*}}=3879.90$
MeV), which are the lowest OZI-allowed decay channels, the $X(3872)$ can be
seen as a quasi-bound state.

Here we will study the $X(3872)$ as a unitarized mesonic state, that
is, one with both quark-antiquark and meson-meson (MM) components.
A previous configuration-space calculation \cite{Coito:2012vf} with
$c\bar{c}$ and $D^{0}D^{0*}$components predicted a state with
approximately $7.5\%$ $c\bar{c}$. We now generalize that calculation
to include other possible channels.

Electromagnetic (EM) decays of the $X(3872)$ were observed by Belle
\cite{Bhardwaj:2011dj}, Babar \cite{Aubert:2008ae} and
LHCb \cite{Aaij:2014ala}. Babar and
LHCb observed decays into $J/\psi\gamma$ and $\psi(2S)\gamma$, and
found the ratio of partial decay widths 
\[
\mathcal{R}_{\psi}=\frac{\Gamma(\psi(2S)\gamma)}{\Gamma(J/\psi\gamma)}
\]
to be of the order of 2.5--3.5, whereas Belle did not observe the decay
into $\psi(2S)\gamma$ at all and set an upper limit on the value of
$\mathcal{R}_{\psi}$ (see Table \ref{Table:emdecayexp}).
\begin{table}[t]
\begin{centering}
\begin{tabular}{|c|c|}
\hline 
Collaboration & $\mathcal{R}_{\psi}$\tabularnewline
\hline 
\hline 
Belle \cite{Bhardwaj:2011dj} & $<2.1$\tabularnewline
\hline 
BaBar \cite{Aubert:2008ae} & $3.4\pm1.4$\tabularnewline
\hline 
LHCb \cite{Aaij:2014ala} & $2.46\pm0.64\pm0.29$\tabularnewline
\hline 
\end{tabular}
\par\end{centering}
\protect\caption{Measured values of the EM rate ratio $\mathcal{R}_{\psi}$.}
\label{Table:emdecayexp}
\end{table}

\section{Method}
\label{method}
We first derive the wave functions of $J/\psi$, $\psi(2S)$ and $X(3872)$,
considering all $c\bar{c}$, $DD$ (only for vector charmonia), $DD^{*}$ and
$D^{*}D^{*}$ channels, where $D^{(*)}$ is shorthand for $D^{(*)0}$,
$D^{(*)\pm}$, or $D_s^{(*)\pm}$. With these, the EM transition matrix elements
and resulting decay widths will be calculated. 

In the present model, a unitarized meson is not just a $q\bar{q}$ state
but it also has MM components:
\begin{equation}
|\psi\rangle=\sum_{c}|\psi_{q\bar{q}}^{c}\rangle+\sum_{j}|\psi_{MM}^{j}\rangle
\; .
\end{equation}
In the quark-antiquark sector we have confinement realized
through a harmonic-oscillator (HO) potential with universal (i.e.,
mass-independent) frequency:
\begin{equation}
V_{Q\bar{Q}}(r)=\frac{1}{2}\mu_{c}\omega^{2}r^{2} \; .
\end{equation}
As for the MM sector, we assume no direct interactions and only
a string-breaking potential that links the $q\bar{q}$ and MM channels
to one another:
\begin{equation}
V_{cj}=\frac{\lambda g_{cj}}{2\mu_{c}}\delta(r-a) \; .
\end{equation}
We take the parameters $m_{c}=1.562\,\mbox{GeV}$ and $\omega=0.190\,\mbox{GeV}$
unchanged with respect to all our previous work.
In the $1^{--}$ and $1^{++}$ cases, somewhat different values of the
overall coupling $\lambda$ will be applied, viz.\ $\lambda_{\psi}$ and
$\lambda_{X}$, respectively, to be determined from the physical charmonium
masses. Furthermore, the $J/\psi$ and $\psi(2S)$ masses will also be used to
fix the value of the string-breaking distance $a$, which we will take the same
for the $X(3872)$. Finally, the $g_{cj}$ are $^{3\!}P_{0}$ coupling
coefficients.

Next we solve the coupled-channel Schr\"{o}dinger equation
\begin{equation}
\left[\begin{array}{cc}
\hat{h}_{q\bar{q}}^{c} & V_{cj}\\
V_{jc}^{\dagger} & \hat{h}_{MM}^{j}
\end{array}\right]\left[\begin{array}{c}
u_{c}\\
v_{j}
\end{array}\right]=E\left[\begin{array}{c}
u_{c}\\
v_{j}
\end{array}\right] \; ,
\end{equation}

with 
\begin{eqnarray*}
\hat{h}_{q\bar{q}}^{c} & = & m_{q}^{c}+m_{\bar{q}}^{c}+\frac{\hbar^{2}}{2\mu_{c}}\big(-\frac{d^{2}}{dr^{2}}+\frac{l_{c}(l_{c}+1)}{r^{2}}\big)+\frac{1}{2}\mu_{c}\omega^{2}r^{2} \; , \\
\hat{h}_{MM}^{j} & = & M_{1}^{j}+M_{2}^{j}+\frac{\hbar^{2}}{2\mu_{j}}\big(-\frac{d^{2}}{dr^{2}}+\frac{L_{j}(L_{j}+1)}{r^{2}}\big) \; .
\end{eqnarray*}
The solutions are known for $r\neq a$ and appropriate boundary conditions:
\begin{equation}
u_{c}(r)=\left\{\begin{array}{r}
\!\!\!a_{c}M(-\nu_{c},l_{c}+\frac{3}{2},\mu_{c}\omega r^{2})e^{-\frac{1}{2}\mu_{c}\omega r^{2}}r^{1+l_{c}} \; , \;\;\; r<a \; , \\
\!\!\!b_{c}U(-\nu_{c},l_{c}+\frac{3}{2},\mu_{c}\omega r^{2})e^{-\frac{1}{2}\mu_{c}\omega r^{2}}r^{1+l_{c}} \; , \;\;\; r>a \; ,
\end{array}\right.
\end{equation}
and
\begin{equation}
v_{j}(r)=\left\{\begin{array}{r}
\!\!\!A_{j}i_{L_{j}}(q_{j}r)\, r \; , \;\;\; r<a \; , \\
\!\!\!B_{j}k_{L_{j}}(q_{j}r)\, r \; , \;\;\; r>a \; .
\end{array}\right.
\end{equation}
Using now continuity of the wave function and discontinuity of its derivative,
we can solve the equations for $a_{c}$, $b_{c}$, $A_{j}$ and $B_{j}$.
The value of the energy $E$ (for fixed $\lambda$) or coupling $\lambda$
(for fixed $E$) is then given by the equation ($\alpha_{c}\equiv a_{c}M_{c}$)
\begin{align}
\Big(\frac{U_{c}'}{U_{c}}-\frac{M_{c}'}{M_{c}}\Big)\alpha_{c} & =\lambda^{2}\sum_{jd}\frac{\mu_{j}g_{cj}}{2\mu_{c}\omega q_{j}a^{3}}\Big(\frac{k{}_{j}'}{k_{j}}-\frac{i_{j}'}{i_{j}}\Big)^{-1}\:\frac{e^{\frac{1}{2}(\mu_{c}-\mu_{d})\omega a^{2}}g_{dj}}{\mu_{d}}\,\alpha_{d} \; .
\end{align}

For more details, see \cite{Cardoso:2014xxx}.

\section{Wave functions}

\begin{figure}[t]
\centering{}\includegraphics[width=0.4\columnwidth]{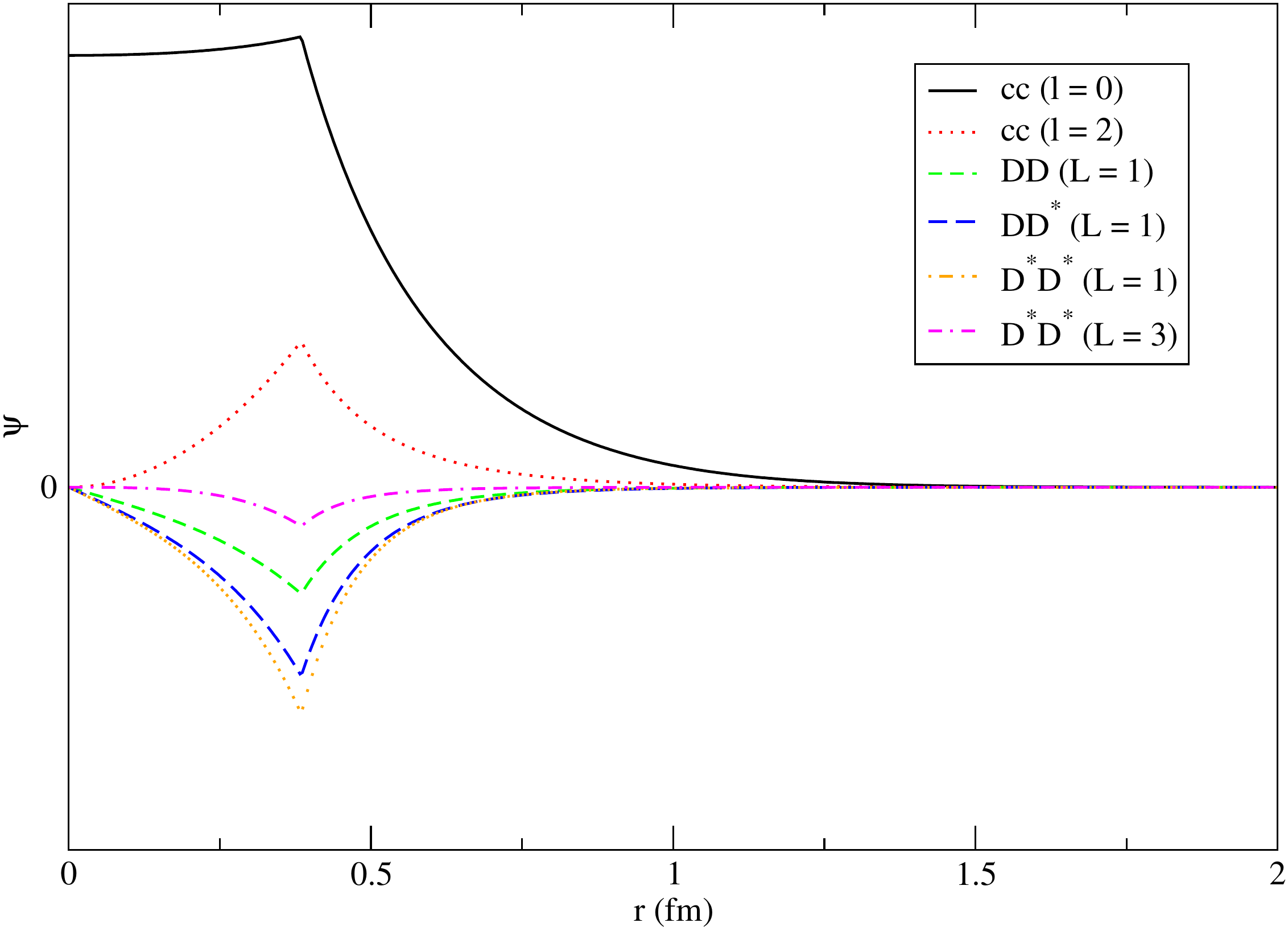}\hspace{1.0cm}\includegraphics[width=0.4\columnwidth]{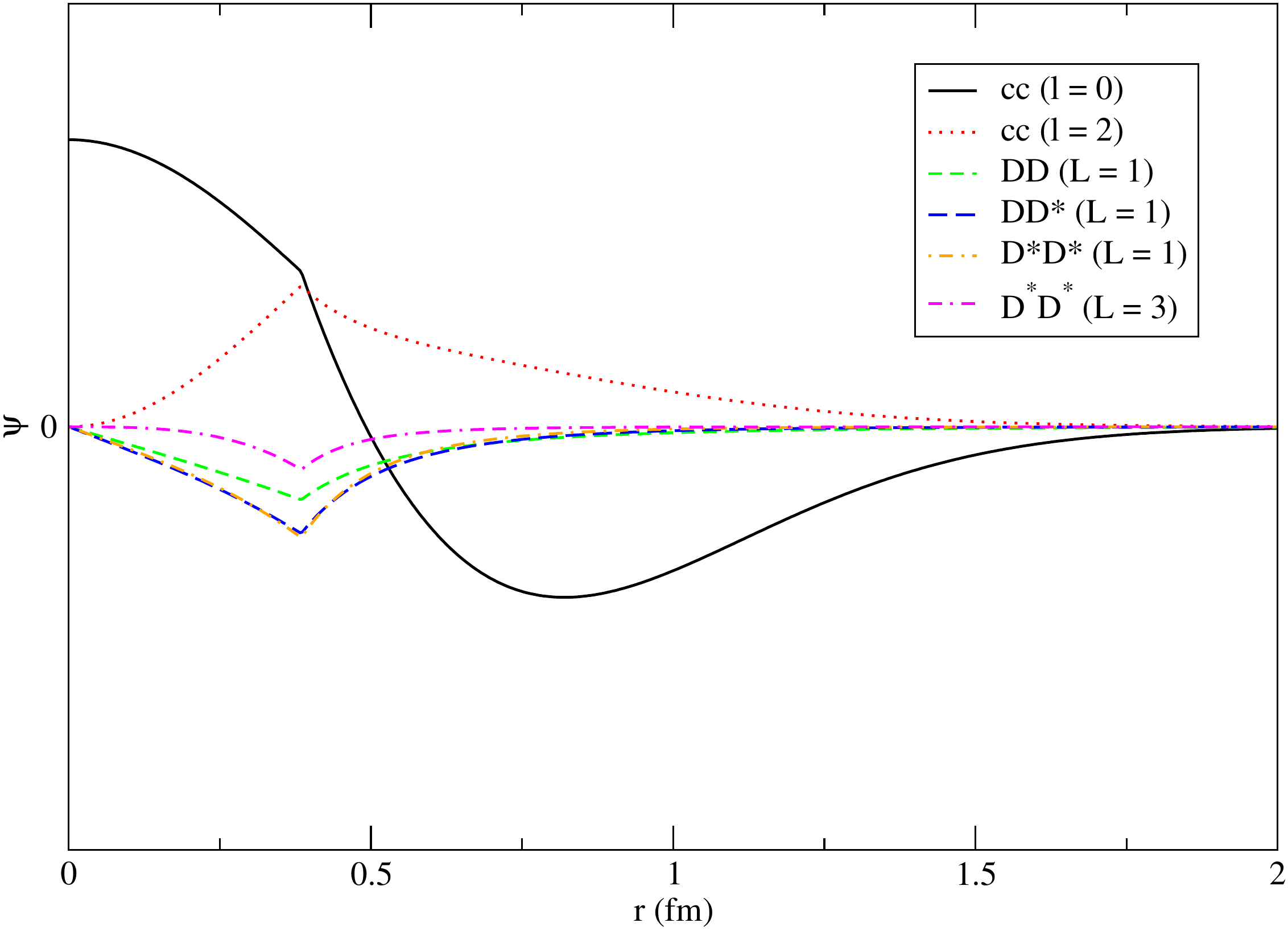}\protect\caption{Wave-function components of the $J/\psi$ and $\psi(2S)$}
\label{Fig:vector_mesons}
\end{figure}

Using the method outlined in Sec.~\ref{method}, and fitting $\lambda_\psi$
as well as $a$ to the experimental $J/\psi$ and $\psi(2S)$ masses, we find
$\lambda_\psi=2.53$ and $a=1.95$~GeV$^{-1}$. The resulting
wave-function components are plotted in Fig.~\ref{Fig:vector_mesons}. Next
\begin{figure}[t]
\centering{}\includegraphics[width=0.5\columnwidth]{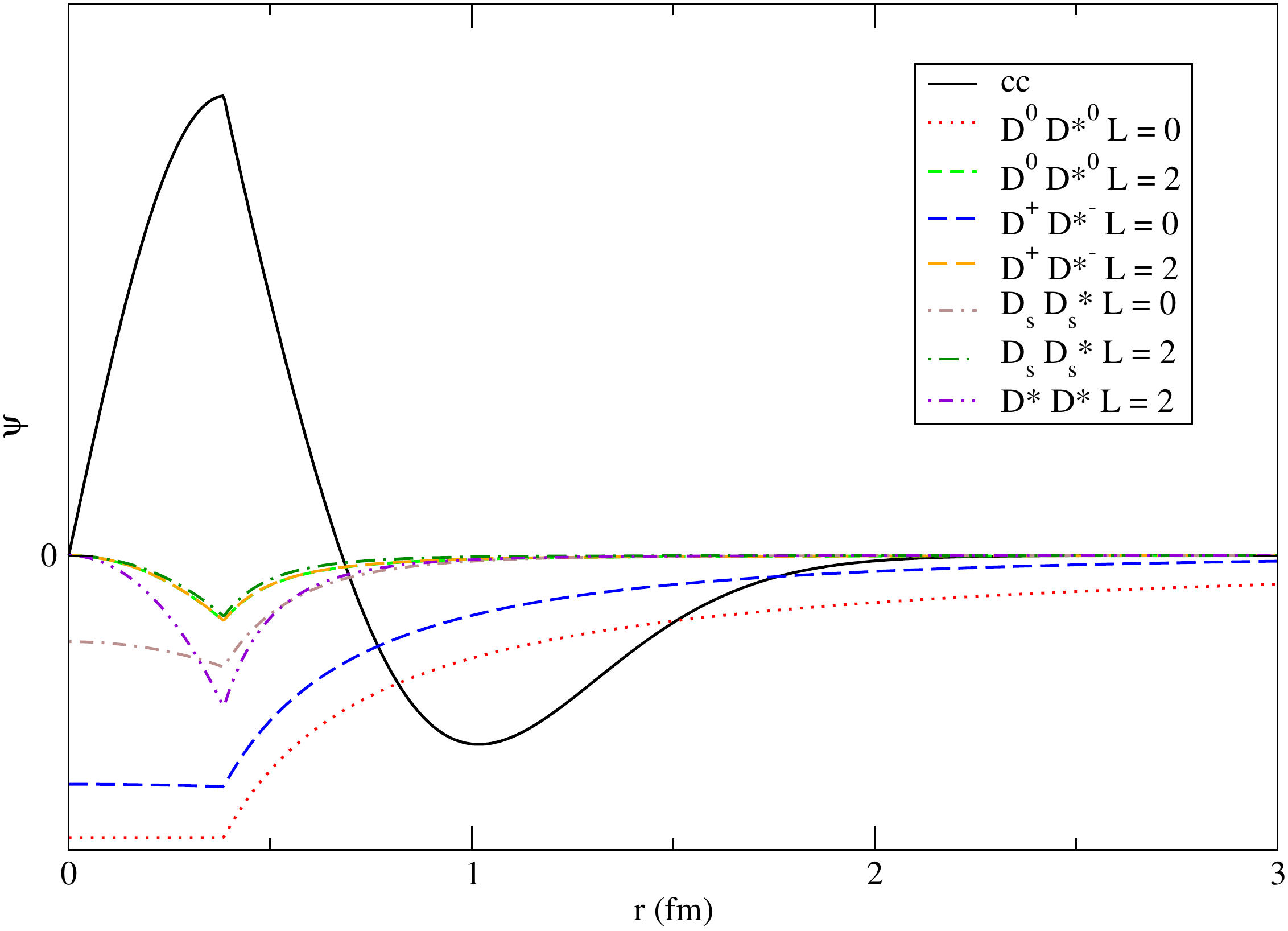}\protect\caption{Wave-function components of the $X(3872)$}
\label{Fig:Xwavefunction}
\end{figure}
we adjust $\lambda_X$ to the $X(3872)$ mass while keeping $a$ the same,
which yields the wave-function components shown in
Fig.~\ref{Fig:Xwavefunction}. The three wave-function compositions 
are given in Table~\ref{Table:comp}. We see that
\begin{table}[b]
\centering{}%
\begin{tabular}{|c|c|c|c|c|c|}
\hline 
 & $c\bar{c}$ & $DD$ & $D^0D^{*0}$ & $D^\pm D^{*\mp}$ & $D^{*}D^{*}$\tabularnewline
\hline 
\hline 
$J/\psi$ & 83.6\% & 2.1\% & \multicolumn{2}{c|}{6.0\%} & 8.3\%\tabularnewline
\hline 
$\psi(2S)$ & 94.5\% & 1.3\% & \multicolumn{2}{c|}{2.1\%} & 2.1\%\tabularnewline
\hline 
$X(3872)$ & 26.8\% & - & 65.0\% & 7.0\% & 1.2\%\tabularnewline
\hline 
\end{tabular}\protect\caption{Compositions of the three charmonia 
($D^{(*)}$: shorthand, see text.)}
\label{Table:comp}
\end{table}
the $J/\psi$ and $\psi(2S)$ are mostly $c\bar{c}$
states, whereas the $X(3872)$ has a dominant $D^0D^{*0}$ component.
Still, its $c\bar{c}$ probability of 26.8\% is a huge increase as compared to
the 7.5\% in \cite{Coito:2012vf}.

\section{Electromagnetic decay}

To compute the EM decay widths we use the Fermi golden
rule 
\begin{equation}
\Gamma_{i\rightarrow f}=\frac{2\pi}{\hbar}|\langle\Psi_{f}|\hat{H}_{int}|\Psi_{i}\rangle|\rho_{f} \; ,
\label{eq:fermigolden}
\end{equation}
with density of states $\rho_{f}=\frac{1}{2\pi\hbar c}$
\cite{Verschuren:1991bg}.
To evaluate the matrix elements in \ref{eq:fermigolden}, we note that
the initial and final states are given by $|\Psi_{i}\rangle=|\psi_{nJM}\rangle\otimes|0\rangle$
and $|\Psi_{f}\rangle=|\psi_{n'J'M'}\rangle\otimes|\gamma_{\lambda klm}\rangle$,
where $l$ and $m$ are the angular-momentum quantum numbers, and $\lambda$
the polarization.

Expanding the wave function, we get a matrix element
\begin{equation}
\langle\Psi_{f}|\hat{H}_{int}|\Psi_{i}\rangle=\sum_{cc'}\langle\psi_{q\bar{q}}^{c}|\hat{h}_{int}^{cc'}|\psi_{q\bar{q}}^{c'}\rangle+\sum_{jj'}\langle\psi_{MM}^{j}|\hat{h}_{int}^{jj'}|\psi_{MM}^{j'}\rangle \; ,
\end{equation}
as we only consider transitions of the types $(Q\bar{Q})^{*}\rightarrow Q\bar{Q}+\gamma$
and $(M_{1}M_{2})^{*}\rightarrow M_{1}M_{2}+\gamma$, neglecting those
like $M_{1}^{*}M_{2}^{*}\rightarrow M_{1}M_{2}+\gamma$.

The interaction Hamiltonian $\hat{h}_{int}$ is obtained from minimal
coupling, accounting for a possible anomalous magnetic moment. In
the radiation gauge $\nabla.\mathbf{A}=0$ and $A^{0}=0$, and neglecting
the $\mathbf{A}^{2}$ term, we have
\begin{equation}
\hat{h}_{int}=\sum_{i}\frac{iQ_{i}}{m_{i}c}\,\mathbf{A}(\mathbf{x}_{i})\cdot\nabla_{i}-\mu_{i}\mathbf{S}_{i}\cdot\mathbf{B}(\mathbf{x}_{i}) \; .
\end{equation}
The EM vector potential is expanded as
\[
\mathbf{A}(\mathbf{r},t)=\sqrt{4\pi}\hbar c\sum_{\lambda lm}\int\frac{dk}{2\pi}\frac{1}{\sqrt{2\omega_{k}}}\big[\mathbf{f}_{klm}^{(\lambda)}(\mathbf{r})e^{-i\omega_{k}t}a_{\lambda lm}(k)+h.\, c.\big] \; ,
\]
with $a_{\lambda lm}$ being photon-annihilation operators. Components
with $\lambda=e$ correspond to electric multipole radiation and
the ones with $\lambda=m$ to magnetic multipole radiation. For the
same $l$, they have opposite parity.

The $X(3872)$ ($1^{++}$ state) can only decay decay into $J/\psi$
and $\psi(2S)$ ($1^{--}$ states) by emitting electric-dipole ($l=1$)
or magnetic-quadrupole ($l=2$) photons.

The computation of the matrix elements is carried out as in \cite{Verschuren:1991bg}.
The resulting EM decay widths are presented in Table~\ref{Table:widths}.
\begin{table}[ht]
\begin{centering}
\begin{tabular}{|c|c|c|c|c|}
\hline 
 & Complete & $c\bar{c}$ & MM & Quenched\tabularnewline
\hline 
\hline 
$\Gamma_{e}(X\rightarrow J/\psi\gamma)$ & 24.2 & 14.9  & 1.11 & 0.48\tabularnewline
\hline 
$\Gamma_{m}(X\rightarrow J\psi\gamma)$ & 0.44 & 0.34 & 0.01 & 0.14\tabularnewline
\hline 
$\Gamma_{e}(X\rightarrow\psi'\gamma)$ & 28.8 & 28.0 & 0.01 & 158\tabularnewline
\hline 
$\Gamma_{m}(X\rightarrow\psi'\gamma)$ & 0.07 & 0.07 & 0.00 & 0.26\tabularnewline
\hline 
\end{tabular}
\par\end{centering}
\protect\caption{Computed EM decay widths in keV. The second and
third columns show the hypothetical widths from the $c\bar{c}$
and $MM$ components only. The last column gives the predictions of
an HO quenched quark model, with the same $m_{c}$
and $\omega$ as in the unquenched case. Note that these numbers are slighty
different from those presented at the workshop, after correction of minor
numerical errors.}
\label{Table:widths}
\end{table}
We obtain an EM rate ratio $\mathcal{R}_{\psi}=1.17$.

\section{Conclusions}

We have generalized a previous configuration-space calculation
\cite{Coito:2012vf} of the $X(3872)$ by
including more MM channels. Thus we obtained an increase of the
total $c\bar{c}$ probability from $7.5\%$ to $26.8\%$. This seemingly
paradoxical result has a simple explanation: the inclusion of more
MM channels leads to a reduction of the $D^{0}D^{*0}$ component,
which --- due to its long tail --- was responsible for an MM probability
exceeding 90\% \cite{Coito:2012vf}. Table~\ref{Table:widths} shows that
unquenching very strongly affects the EM widths.
Our prediction of the ratio $\mathcal{R}_{\psi}=1.17$ is consistent
with the result of Belle, but does not fully agree with BaBar and
LHCb. However, there is an enormous improvement when compared to a quenched
HO calculation. For a more detailed discussion, see
\cite{Cardoso:2014xxx}.

M.~Cardoso was supported by FCT, contract SFRH/BPD/73140/2010. 

\bibliographystyle{unsrtnt}
\bibliography{bib}

\begin{thebibliography}{10}

\bibitem{Choi:2003ue}
S.-K.~Choi {\em et~al.} \/[Belle Collaboration],
{\em Phys.\ Rev.\ Lett.} \/{\bf91}, 262001 (2003).

\bibitem{Acosta:2003zx}
D.~Acosta {\em et~al.} \/[CDF Collaboration],
{\em Phys.\ Rev.\ Lett.} \/{\bf93}, 072001 (2004).

\bibitem{Abazov:2004kp}
V.~M.~Abazov {\em et~al.} \/[D0 Collaboration],
{\em Phys.\ Rev.\ Lett.} \/{\bf93}, 162002 (2004).

\bibitem{Agashe:2014kda}
K.~A.~Olive {\em et~al.} \/[Particle Data Group Collaboration],
{\em Chin.\ Phys.} \/{\bf C38}, 090001 (2014).

\bibitem{Aaij:2013zoa}
R.~Aaij {\em et~al.} \/[LHCb Collaboration], 
{\em Phys.\ Rev.\ Lett.} \/{\bf110}, 222001 (2013).

\bibitem{Aubert:2004zr}
B.~Aubert {\em et~al.} \/[BaBar Collaboration],
{\em Phys.\ Rev.} \/{\bf D71}, 031501 (2005).

\bibitem{Choi:2011fc}
S.-K. Choi {\em et~al.} \/[Belle Collaboration],
{\em Phys.\ Rev.} \/{\bf D84}, 052004 (2011).

\bibitem{Coito:2012vf}
S.~Coito, G.~Rupp, E.~van Beveren,
{\em Eur.\ Phys.\ J.} \/{\bf C73}, 2351 (2013).

\bibitem{Bhardwaj:2011dj}
V.~Bhardwaj {\em et~al.} \/[Belle Collaboration],
{\em Phys.\ Rev.\ Lett.} \/{\bf107}, 091803 (2011).

\bibitem{Aubert:2008ae}
B.~Aubert {\em et~al.} \/[BaBar Collaboration],
{\em Phys.\ Rev.\ Lett.} \/{\bf102}, 132001 (2009).

\bibitem{Aaij:2014ala}
R.~Aaij {\em et~al.} \/[LHCb Collaboration],
{\em Nucl.\ Phys.} \/{\bf B886}, 665 (2014).

\bibitem{Cardoso:2014xxx}
M.~Cardoso, G.~Rupp, E.~van Beveren, 
arXiv:1411.1654v2 [hep-ph], accepted for publication in
{\it Eur.\ Phys.\ J.\ C}.

\bibitem{Verschuren:1991bg}
A.~G.~M.~Verschuren, C.~Dullemond, E.~van Beveren,
{\em Phys.\ Rev.} \/{\bf D44}, 2803 (1991).

\end{thebibliography}

\end{document}